\documentclass[twocolumn,showpacs,preprintnumbers,amsmath,amssymb]{revtex4}
\usepackage{graphicx}% Include figure files
\usepackage{dcolumn}% Align table columns on decimal point
\usepackage{bm}% bold math
\usepackage{psfig}
\usepackage{epsfig}

\begin{document}

\newcommand{\vAi}{{\cal A}_{i_1\cdots i_n}}
\newcommand{\vAim}{{\cal A}_{i_1\cdots i_{n-1}}}
\newcommand{\vAbi}{\bar{\cal A}^{i_1\cdots i_n}}
\newcommand{\vAbim}{\bar{\cal A}^{i_1\cdots i_{n-1}}}
\newcommand{\htS}{\hat{S}}
\newcommand{\htR}{\hat{R}}
\newcommand{\htI}{\hat{I}}
\newcommand{\htB}{\hat{B}}
\newcommand{\htD}{\hat{D}}
\newcommand{\htV}{\hat{V}}
\newcommand{\cT}{{\cal T}}
\newcommand{\cM}{{\cal M}}
\newcommand{\cMs}{{\cal M}^*}
\newcommand{\vk}{{\bf k}}
\newcommand{\vK}{{\\bf K}}
\newcommand{\vb}{{\textstyle{\bf b}}}
\newcommand{{\vp}}{{\vec p}}
\newcommand{{\vq}}{{\vec q}}
\newcommand{\vQ}{{\vec Q}}
\newcommand{\vx}{{\textstyle{\bf x}}}
\newcommand{\tr}{{{\rm Tr}}}
\newcommand{\beq}{\begin{equation}}
\newcommand{\eeq}[1]{\label{#1} \end{equation}}
\newcommand{\half}{{\textstyle \frac{1}{2} }}
\newcommand{\lton}{\mathrel{\lower.9ex \hbox{$\stackrel{\displaystyle
<}{\sim}$}}}
\newcommand{\gton}{\mathrel{\lower.9ex \hbox{$\stackrel{\displaystyle
>}{\sim}$}}}
\newcommand{\ee}{\end{equation}}
\newcommand{\ben}{\begin{enumerate}}
\newcommand{\een}{\end{enumerate}}
\newcommand{\bit}{\begin{itemize}}
\newcommand{\eit}{\end{itemize}}
\newcommand{\bc}{\begin{center}}
\newcommand{\ec}{\end{center}}
\newcommand{\bea}{\begin{eqnarray}}
\newcommand{\eea}{\end{eqnarray}}
\newcommand{\beqar}{\begin{eqnarray}}
\newcommand{\eeqar}[1]{\label{#1}\end{eqnarray}}
\newcommand{\bra}[1]{\langle {#1}|}
\newcommand{\ket}[1]{|{#1}\rangle}
\newcommand{\norm}[2]{\langle{#1}|{#2}\rangle}
\newcommand{\brac}[3]{\langle{#1}|{#2}|{#3}\rangle}
\newcommand{\hilb}{{\cal H}}
\newcommand{\pleft}{\stackrel{\leftarrow}{\partial}}
\newcommand{\pright}{\stackrel{\rightarrow}{\partial}}

\begin{flushright}
%DRAFT 4/7 10pm
%\vskip .5cm
\end{flushright} \vspace{1cm}

\title{3D Jet Tomography of Twisted Strongly Coupled Quark Gluon Plasmas
}

\author{A.~Adil}%
\email{azfar@phys.columbia.edu}

\author{M.~Gyulassy}
\email{gyulassy@phys.columbia.edu}

\affiliation{Columbia University, Department of
Physics, 538 West 120-th Street, New York, NY 10027
}%

\date{\today}% It is always \today, today,
             %  but any date may be explicitly specified

\begin{abstract}
  The {\em triangular} enhancement of the rapidity distribution
  of hadrons produced in p+A reactions relative to p+p is a leading
  order in ($A^{1/3}/\log s$) violation of longitudinal boost
  invariance at high energies.  
In $A+A$ reactions this leads
to a  {\em
    trapezoidal}
  enhancement of the local rapidity density of produced gluons.
The local rapidity gradient is proportional to the
  local participant number asymmetry, and leads to an effective
  rotation  in the %
  reaction plane. We propose that three dimensional jet tomography ,
  correlating the long range rapidity and azimuthal dependences of the
  nuclear modification factor, $R_{AA}(\eta,\phi,p_\perp; b>0)$,
 can be used to
  look for this intrinsic longitudinal
boost violating structure of $A+A$ collisions to image the
  produced twisted strongly coupled quark gluon plasma (sQGP).
 In addition to dipole and elliptic azimuthal moments of
  $R_{AA}$, a significant high $p_\perp$ octupole  moment
  is predicted away from midrapidity. The azimuthal angles of maximal
  opacity and hence minima of $R_{AA}$ are rotated
  away from the normal to the reaction plane by an `Octupole Twist''
  angle, $\theta_3(\eta)$, at forward rapidities.
 \end{abstract}

\pacs{12.38.Mh; 24.85.+p; 25.75.-q}

\maketitle

%%%%%%%%%%%%%%%%%%%%%%%%%%%%%%%%%%%%%%%%%%%%%%%%%%%%%%%%%%%%%%%%%%%%%%%%%%
\section{Introduction}

The experimental discovery\cite{Adcox:2004mh} of
a new form of strongly interacting Quark Gluon Plasma matter (called the 
sQGP\cite{Gyulassy:2004zy}) at the
Relativistic Heavy Ion Collider (RHIC) has raised many
new questions about the physics
of ultra-dense matter produced in relativistic nuclear collisions.
Striking new phenomena
observed thus far include jet quenching, fine structure of elliptic
flow, nuclear gluon saturation, and the baryon-meson
anomaly. The primary experimental parameters
available at the RHIC
 are the beam energy, $\sqrt{s}=20-200$ AGeV, and the
variation of nuclear geometry through different $B+A$ combinations.
Variations of the global 
nuclear geometry via impact parameter or centrality cuts
adds another set of
important experimental methods and instruments with which the geometric
configuration of the sQGP can be systematically varied.
 In non central
collisions of nuclei, the transverse area of the participant 
interaction region has an elliptic shape with an aspect ratio
or eccentricity that can be varied in a controlled way via 
global multiplicity cuts.
In this paper we explore a more subtle aspects
of the 3D geometry of sQGP produced in non central A+A collisions
that can be explored by extending jet tomography
techniques\cite{Gyulassy:2003mc} to full 3D taking the third ``slice''
dimension to be the rapidity variable.

Strong elliptic transverse flow resulting from transverse
elliptic asymmetry provides the key barometric probe of the sQGP
equation of state.  A common simplifying assumption in most
hydrodynamic and transport analyses of the data thus far is that the initial
conditions are longitudinally boost invariant \cite{Bjorken:1982qr}.
However, the rapidity dependence of elliptic flow data
clearly demonstrate important
deviations of the global collective 
dynamics from boost invariance away from the
midrapidity region\cite{phobos_v2PRL,starv2}.  Even the most advanced 3+1D
hydrodynamic calculations\cite{Hirano:2001eu,
  Hirano:2004rs,Magas:2002ge}, which relax the ideal 2+1D Bjorken
initial conditions, have not yet been able to reproduce the observed
rapidity dependence of elliptic flow.  The solution may require 
better understanding of the geometry of the initial sQGP in addition
to better theoretical control over the growing 
dissipative hadronic corona
at high rapidities\cite{Gyulassy:2004zy}. 

As we show below, the dimensionless parameter that controls violations
of longitudinal boost invariance in $A+A$ reactions at finite energies
is the {\em local} relative rapidity slope
of the low transverse momentum partons
 \beq
\delta=\frac{A^{1/3}-1}{2 Y} 
\; , \eeq{delta}
where $2 Y=\log s$ is the rapidity gap between
the projectile and target. 
In $Au+Au$ at RHIC  $\sqrt{s}=200$ AGeV, $\delta\approx 0.45$
cannot be ignored.
Even in $Pb+Pb$ at the Large Hadron Collider (LHC) with $\sqrt{s}=5500$ AGeV, $\delta\approx 0.28$
will not be much smaller. 

In this paper, 
we explore some observable 
consequences of such
violations of
longitudinal boost invariance. We predict that there is
an intrinsic rotation of the initially produced matter in the
reaction plane in non-central $A+A$ that can induce an octupole moment
of the nuclear modification factor
$R_{AA}(\eta\ne0,\phi,p_\perp; b>0)$ at moderate to high $p_\perp$ 
away from mid-rapidities.
This effect arises from the combination of
two basic properties of high energy
hadron-nucleus reactions with jet quenching dynamics in $A+A$. 

One important property is that the rapidity density
of produced hadrons is observed\cite{Busza:2004mc} 
to be {\em Triangular} relative to the density produced in
$p+p$ reactions with a slope given by $\sim \delta$.
The second basic property follows from the Eikonal-Glauber
reaction geometry that predicts {\em local} variations of 
the projectile and target participant number density 
in the transverse plane, $\vx_\perp$.
We propose that
the long range correlations of the rapidity
and azimuthal dependence of jet quenching can be used to extend current
2D $R_{AA}(\phi)$ jet tomography\cite{Gyulassy:2003mc} 
to ``image'' the 3D geometry
of the twisted sQGP initial state at forward rapidities $|\eta|\sim 1-3 $.

In the next section we begin with a review
of the $p+A$ triangle and its implications for
$A+A$ initial conditions. In section \ref{number1} 
we discuss the role of more realistic diffuse nuclear geometry 
 on  the local participant
and binary collision
  densities. In Section \ref{number2} we 
  calculate jet opacities as a function of $\eta,\phi$. In Section
  \ref{number3} jet attenuation
through the tilted bulk sQGP density in the reaction plane 
is studied, and in 
 Section \ref{number4} the azimuthal Fourier moments
  of the nuclear modification factor are computed
as a function of rapidity.
Our proposed new observable,
  the Octupole Twist angle $\theta_3(\eta)$, 
is computed.
We conclude in Section \ref{number5}.

\section{The P+A Triangle}
In p+A reactions, the ratio
$R_{pA}(\eta)=(dN_{pA}/d\eta)/(dN_{pp}/d\eta)$ of produced
low $p_\perp$ hadrons as a function of rapidity, $\eta$,
 was found at all energies to be  ``triangular'' with
the height near the rapidity $(-Y)$ of the target nucleus proportional to 
$\nu_A \propto A^{1/3}$, the average number of ``wounded target nucleons''
\cite{
WNM,Busza:2004mc}. (Note that we are using the variable $\eta$ to denote the particle rapidity rather than the particle pseudorapidity, which $\eta$ is conventionally linked to.  This is to avoid later confusion with the variable $y$, used to denote the transverse vertical coordinate in this paper.)
Excess nuclear enhancement $\propto A^{1/3+q}$ near the target rapidity is due
to intra-nuclear cascading of low relative energy hadrons produced in the
target fragmentation region.  We will not consider this extra nuclear modification further in this paper.

The projectile proton in the center of rapidity frame has 
rapidity, $Y$, and
interacts at a transverse impact parameter, $\vb$, on the average with
 \beq
\nu_A(\vb)\approx \sigma_{in}T_A(\vb)=\sigma_{in}\int dz \rho_A(z,\vb)\propto
A^{1/3}
\eeq{nut} 
target nucleons with rapidity $-Y$. Here
$\sigma_{in}(\sqrt{s})$ is the NN inelastic cross section, and the
Glauber profile function, $T_A(\vb)$ measures the number target nucleons
per unit area of nucleus with proper density $\rho_A(r)$.  

The simplest way to understand the origin of the pA triangle
is in terms of phenomenological string models of high
energy inelastic interactions. Each wounded nucleon\cite{WNM} in
both projectile and target is diffractively
excited into high mass $\sim \sqrt{s}/2$ states
that decay as beam jets into relatively low transverse 
momentum ($p_T \sim \Lambda_{QCD}$) 
over a wide rapidity interval. Wounded target nucleons decay over an
interval $-Y<\eta \lton 0 $ 
while projectile nucleons decay over $0\gton\eta< Y$.
The resulting asymmetry 
of the produced hadron rapidity density $dN_{pA}/d\eta$
thus grows approximately linearly with $\nu_A$.

\begin{figure}
\centering
 \epsfig{file=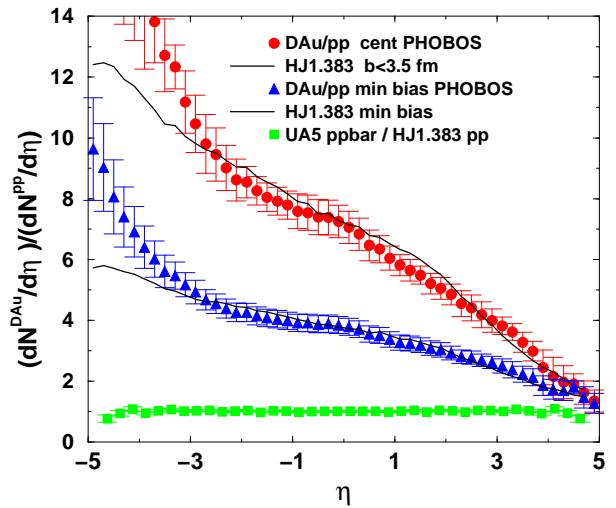,width=2.7in,angle=-90}
  \caption{The approximate triangular form of the ratio of 
the observed $D+Au$\protect{\cite{Back:2004mr}} 
    charged particle pseudorapidity distributions 
to the dashed HIJING $p+p$ curve from Fig.1 are shown for two centrality cuts.
The solid curves are predictions using HIJING 1.383
    \protect{\cite{Wang:1991ht}} for $D+Au/p+p$.
    The ratio of UA5
$p+\bar{p}$\cite{Alner:1986xu} data to the baseline
HIJING p+p
    predictions are shown by filled squares. (In color online) }
  \label{ratdau}
\end{figure}

\begin{figure}
\centering
 \epsfig{file=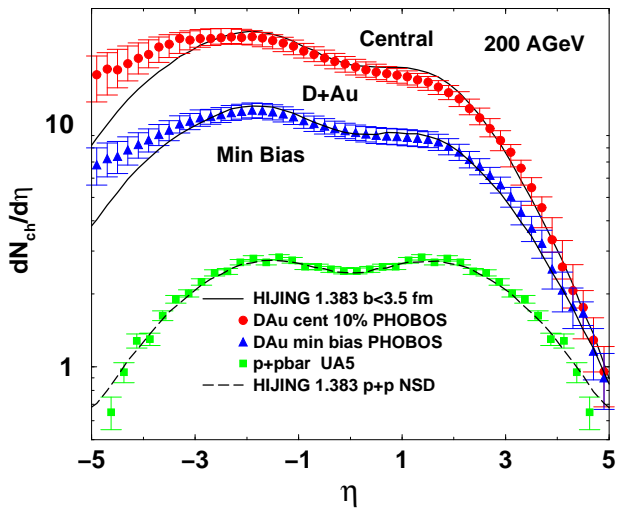,width=2.7in,angle=-90}
  \caption{Asymmetric pseudorapidity distributions of charged hadrons
produced 
in $D+Au$ minimum bias and central 0-10$\%$
reactions at 200 AGeV from PHOBOS\protect{\cite{Back:2004mr}} 
are compared to $p+\bar{p}$ data from UA5\protect{\cite{Alner:1986xu}}.
The  curves show predictions using the HIJING v1.383
code\protect{\cite{Wang:1991ht}}. (In color online)
}
  \label{dau}
\end{figure}

At RHIC\cite{Back:2004mr}, the rapidity triangle
was clearly observed in the Deuterium+Au
control experiment at 200 AGeV relative to $pp$ and
$p\bar{p}$\cite{Alner:1986xu} as shown in Fig. \ref{ratdau}.  The absolute agreement between HIJING calculations and the measured multiplicities can be seen in Fig. \ref{dau}.  the disagreement at extreme negative rapidities marks the presence of intranuclear cascading.  Also note that the figures show pseudorapidity distributions rather than rapidity distributions.  These are the same up to a jacobian suppression factor present in all pseudorapidity distributions ($~15\%$) that can be seen in Fig. \ref{dau}.  This jacobian factor cancels in the ratio.

\begin{figure}
\centering
 \epsfig{file=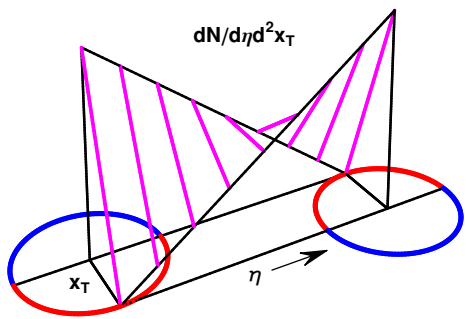,width=2.7in,angle=0}
  \caption{Schematic illustration 
of how local trapezoidal nuclear enhancements
of the rapidity distributions in the 
reaction plane $(x,\eta,y=0)$
twist the bulk initial density about the normal
in non central $A+A$ collisions.
(see eqs.(\protect{\ref{BGK}},\ref{torsan})) (In color online)
}
  \label{dndxt1}
\end{figure}

\begin{figure}
\centering
 \epsfig{file=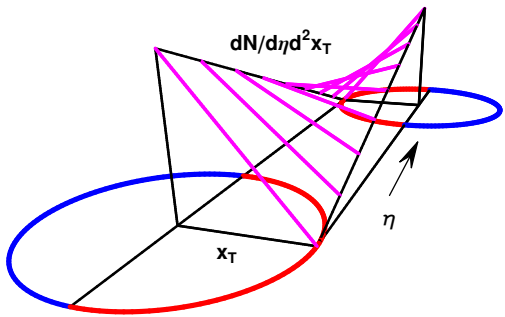,width=2.7in,angle=0}
  \caption{As in Fig.3 from another perspective.
The straight segments in each rapidity slice are
for emphasis only. Realistic diffuse spherical geometries
curve these lines (see Fig.7 below). (In color online)}
  \label{dndxt2}
\end{figure}

One of the early parton model interpretations
 of the rapidity triangle and its
generalization to rapidity ``trapezoids'' in B+A collisions was given in
BGK\cite{Brodsky:1977de}.  The color excitation of wounded baryons
proceeds in this picture with 
a %perturbative QCD 
parton-parton scattering at a rapidity, $y^*$,
distributed approximately uniformly %in the Feynman gas
%model 
between $-Y$ and $Y$. The
color excited projectile nucleon then 
neutralizes by producing a string of hadrons
from $y^* < \eta< Y$ while the target nucleon neutralizes by producing
a string of hadrons from $-Y<\eta <y^*$. The approximately triangular
rapidity density in this model arises naturally from the sum over the
$\nu_A$ target and one projectile fragments. 

BGK\cite{Brodsky:1977de} generalized this model to predict 
``trapezoidal'' distributions that would arise when a row of $\nu_B>1$
projectile nucleons interacts inelastically with a row of $\nu_A>1$
target nucleons. 
This very basic feature of multi-nucleon interactions in
pQCD is at the core of the  phenomenological success of LUND string
models such as FRITIOF\cite{Andersson:1986gw} and
HIJING\cite{Wang:1991ht}.  The curves shown in Figs.1 and 2 are
predictions of the HIJING model\cite{Gyulassy:1994ew} using version  1.383 
that includes the Hulthen distribution of nucleons in Deuterium. 
The magnitude and centrality dependence of the charged particle rapidity 
distributions are well predicted by the basic trapezoidal 
nuclear enhancement of multi-particle production dynamics.
We note that Wang\cite{Wang:2003vy} studied this rapidity
asymmetry also as a function of $p_T$ out
to moderately high $p_\perp$ to test Cronin versus shadowing dynamics
away from midrapidity.

While the {\em global} features of the $p_\perp$ integrated
$dN/d\eta$ are well
accounted for, we focus in this paper 
on the novel {\em local} implications 
of the trapezoidal violations of longitudinal boost invariance
for the initial conditions in $B+A$.
In a non-central collision at an impact parameter
$\vb$, the {\em local} enhancement of bulk particle production relative
to p+p at a particular transverse coordinate
$\vx_\perp$ is approximately given in the BGK model by
\beqar 
\frac{dR_{BA}}{d^2\vx_\perp}(\eta,\vx_\perp;\vb)
&=&\frac{dN^{BA}/d\eta d^2\vx_\perp}{dN^{pp}/d\eta}  \nonumber \\ 
\approx  \nu_A(\vx_\perp-{\vb}/{2}) 
\frac{Y-\eta}{2 Y}&+& \nu_B(\vx_\perp+{\vb}/{2}) \frac{Y+\eta}{2Y} \; \; .
\eeqar{BGK} 
Even if $p+p$ were longitudinally boost invariant with a constant
$dN^{pp}/d\eta$,  the   variation of $\nu_A$ and $\nu_B$
with $\vx_\perp$ leads to local violations 
of longitudinal boost invariance in non-central $B+A$.
The local deviations are controlled  by
\beq
\delta(\vx_\perp;\vb)
= %\frac{\partial R_{BA}}{\partial \eta d^2\vx_\perp}= 
\frac{\nu_B-\nu_A}{2Y} 
\; , \eeq{delta2}
i.e. the local participant asymmetry $\nu_T-\nu_P$ 
at $\vx_\perp$.

For very large energies, $Y\gg 1$,
Bjorken longitudinal boost invariance is restored.
However for Au+Au at RHIC, $Y\approx 5$, $A^{1/3}\sim 6$, and there is
an order unity variation of $\delta$ 
across the transverse profile of the interaction region that
leads to the twisted ribbon geometry
of the local density distribution
as illustrated schematically in Figs. 3,4. In these figures, the coordinate
normal to the reaction plane , $y$, is set to zero, and the
surface $dR_{AA}$ is illustrated as a function
of $x=x_T$ in different $\eta$ slices. In reality
only the fixed $x$ slices
have trapezoidal shape while the $\eta$ slices are rounded
due to the nonlinear dependence of the $\nu_i$
on transverse coordinates.

At the global $dN/d\eta$ level, the integration over the transverse
coordinates leads to
\beqar 
R_{BA}(\eta;\vb)
&=& \frac{dN^{BA}/d\eta}{dN^{pp}/d\eta}  \nonumber \\ 
\approx \half({N_A+ N_B}) &+& \frac{\eta}{2Y}(N_B-N_A)  
\; \; ,
\eeqar{BGKglob} 
where $N_A=\int d^2\vx_\perp \nu_A(\vx_\perp;\vb)$ and $N_B$
are the total number of wounded target and projectile nucleons
interacting at impact parameter $b$. 
What is remarkable about eqs.(\ref{BGK},\ref{BGKglob}) and readily seen on 
in Figs 3,4 is that for {\em symmetric} $A+A$ reactions,
the global ratio may appear to be approximately boost invariant,
but 
the local density retains an {\em intrinsic} rapidity
asymmetry that can be characterized by
a rotation twist in reaction plane
\beq
T_{x\eta} = 
\frac{\partial^2}{\partial \eta \partial x} 
\log\left( \frac{dN}{d\eta d^2\vx_\perp} \right)
\; , \eeq{Tors} 
where $x$ %=\hat{\vb}\cdot \vx_\perp$ 
and 
$y$
%=(\hat{\vb}\times\hat{{\bf \eta}})\cdot\vx_\perp$ 
are the coordinates
in and out of reaction plane. A useful property of this characterization is
that it is independent of the normalization of $dN$ and independent
of the  multiplicative rapidity modulation factor  $dN^{pp}/d \eta$.

For a sharp sphere
geometry with 
\beq
\nu\propto \sqrt{R^2-(x\pm b/2)^2-y^2}
\;  \eeq{sph}
the $x\eta$ twist in the center $x=y=0$ of the oval
interaction region is independent of 
$\eta$ and given by
\beq
T_{x\eta}(\vx_\perp=0) = 
\frac{1}{Y}\frac{2b}{4R^2-b^2}
\;\; . \eeq{torsan}
The twist would of course be absent if nuclei 
were uniformly thick. 
%with constant $\nu$ independent of $\vx_{\perp}$. 
The magnitude for realistic 
spherical nuclei
depends on the actual diffuse Wood Saxon geometry that we
use for our numerical estimate below.
However, (\ref{torsan}) shows that its typical magnitude of $T_{x\eta}$
is $1/(R Y)$ in minimum
bias reactions. 

We also note that the trapezoidal distribution of low $p_T$
multi-particle production in $D+Au$ 
follows more generally 
if the collinear factorized QCD mini-jet dynamics of BGK
is replaced by the 
$k_T$ factorized gluon fusion mechanism in the Color Glass
Condensate (CGC) model\cite{Kharzeev:2002ei,McLerran:1993ni}.
Quantitative differences between these mechanisms
arise at moderate  $p_T > 2-5 GeV$ 
which are currently under 
investigation at RHIC\cite{Adcox:2004mh,Arsene:2004ux}.
They are connected with different physical
approximations to Cronin enhancement
and gluon shadowing\cite{Accardi:2003jh}. 

The empirical trapezoidal
form (\ref{BGK}) in both BGK and CGC cases differs from
local equilibrium initial conditions based on
firetube or firestreak 
models \cite{Hirano:2001eu,Magas:2002ge,Myers:1978bz}.
While such models also predict an effective tilt in the $x\eta$ plane,
those corresponds to possible initial rapidity shear
of the hydrodynamic fluid at fixed $\eta$.
In our approach, there is no initial rapidity shear
and the collective longitudinal 
``fluid'' velocity vanishes in the comoving $(y=\eta)$ frame
over the whole transverse region. However, 
there is an initial 
gradient of the density along the beam axis
that can induce a fluid shear during later evolution.
Polarization has been proposed as an observable sensitive
to rapidity shear\cite{Liang:2004ph}.
We concentrate here on generic trapezoidal (BGK, HIJING, or CGC) 
initial ``twisted sQGP'' conditions because they follow directly
from QCD parton dynamics and well constrained by $p+A$ phenomenology.

\section{Diffuse Nuclear Geometry} \label{number1}

To take the diffuse nuclear geometry into account,
we recall that the local  projectile $B$ and
target $A$ participant densities is given in Glauber
theory by
\bea %
\frac{dN_{\textrm{part}}^{\textrm{B}}}{dxdy}&=& 
T_{B}(r_+) 
(1-e^{-\sigma_{in} T_{A}(r_-)}) \nonumber \\ 
\frac{dN_{\textrm{part}}^{\textrm{A}}}{dxdy}&=& 
T_{A}(r_-) 
(1-e^{-\sigma_{in} T_{B}(r_+)}) \; \; ,
\eeqar{targpart}
where
$r_\pm=\sqrt{(x\pm\frac{b}{2})^{2}+y^{2}}$ and
$\sigma\approx 42$ mb at $\sqrt{s}=200$ AGeV.
Rare jets are produced on the other hand distributed in the transverse plane
according to the the binary collision number distribution 
given by 
\begin{equation}
\frac{dN_{\textrm{Bin}}}{dxdy}=\sigma_H T_{B}(r_+)T_{A}(r_-)
\; \; , \label{nbin}
\end{equation}
where $\sigma_H(p_\perp,\eta)$ is the pQCD jet cross section.
For $B=A$ collisions,
the binary distribution is obviously 
symmetric in the transverse plane at any impact parameter,
unlike the participant distributions. Nuclear shadowing and Cronin
effects can break 
this symmetry, but for the present study we 
assume here that at high enough $p_T$ , these initial state 
nuclear effects can be neglected.

Jet quenching at a given rapidity
in azimuthal direction $\phi$ 
depends on a transverse coordinate line integral
across the sQGP density profile at that rapidity\cite{Gyulassy:2003mc}. 
The initial sQGP
density is assumed here to be proportional to the
local participant density. 
In the BGK model this is then approximately given by
\begin{eqnarray}\label{dNdxdydeta}
\; &\;& \frac{dN_{g}}{dxdyd\eta}=
\frac{C}{2Y} \; \exp [ \frac{ -\eta^{2} }{ \sigma_{\eta}^{2} }]
\; \theta(Y-|\eta|)  \nonumber\\
\; &\;& \left\{ \frac{dN_{\textrm{Part}}^A }{dxdy}
(Y-\eta)%\right.  
+ %\left. 
\frac{dN_{\textrm{Part}}^B }{dxdy}
(Y+\eta) \right\}  \; \; ,
\end{eqnarray}
where we have introduced a Gaussian rapidity envelope
to model $dN^{pp}/d\eta$ at RHIC.
With $Y=5$, $\sigma_{\eta}=3$ and
$C\approx 24$, the $\vx_\perp$ integrated $dN_g/d\eta \; \times 1/3$ 
fits the 
BRAHMS  central $Au+Au\rightarrow \pi^{+},\pi^{-})$ data\cite{brahms_dndy} 
as shown in Fig. 5. 

\begin{figure}
\centering
 \epsfig{file=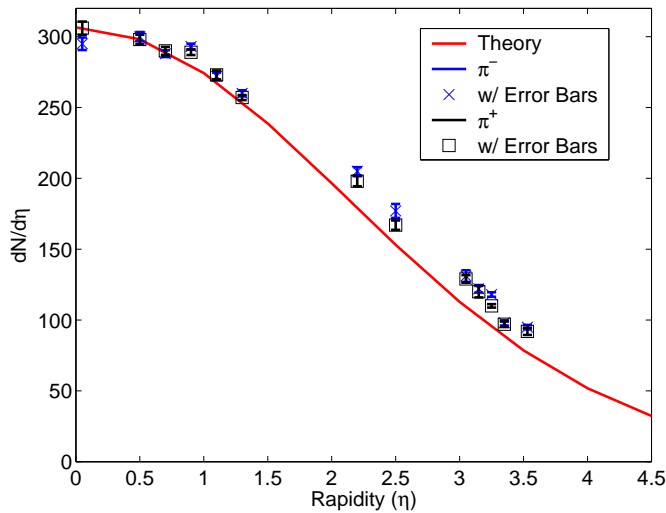,width=3.5in,angle=0}
  \caption{A Gaussian fit of the $\pi^\pm$ 
rapidity density in central $Au+Au$ at 200 AGeV is shown compared to
BRAHMS\protect{\cite{brahms_dndy}} data. (In color online)}
    \label{multcomp}
\end{figure}

\begin{figure}[t!]
\begin{center}
%\hspace*{-.6in}
\psfig{file=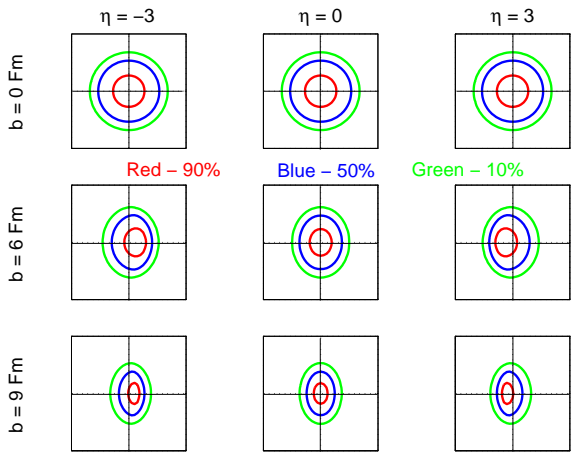,width=3in,angle=0}
\caption{Contours of the twisted sQGP initial density 
in the transverse plane $(x,y)$ in different rapidity
$\eta=-3,0,3$ and  impact parameter
$b=0,6,9$ fm slices. Note the opposite 
transverse shifts at $\eta$ and $-\eta$. (In color online)}
\label{contours}
\end{center}
\end{figure}

\begin{figure}
\centering
 \epsfig{file=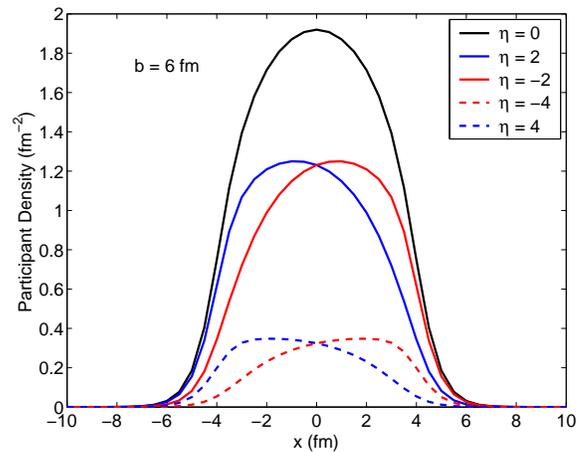,width=3.0in,angle=0}
  \caption{Transverse profile of the twisted sQGP 
at $b=6$ Fm as a function 
of transverse coordinate $x$ at
$y=0$ for $\eta=\pm 2,\pm4$.
The normalization is to the local transverse 
participant density fm$^{-2}$.
 The overall decrease of the 
density away from midrapidity
is due to the approximate Gaussian envelope in Fig.5. (In color online)}
    \label{partdensonline}
\end{figure}

\begin{figure}
\centering
 \epsfig{file=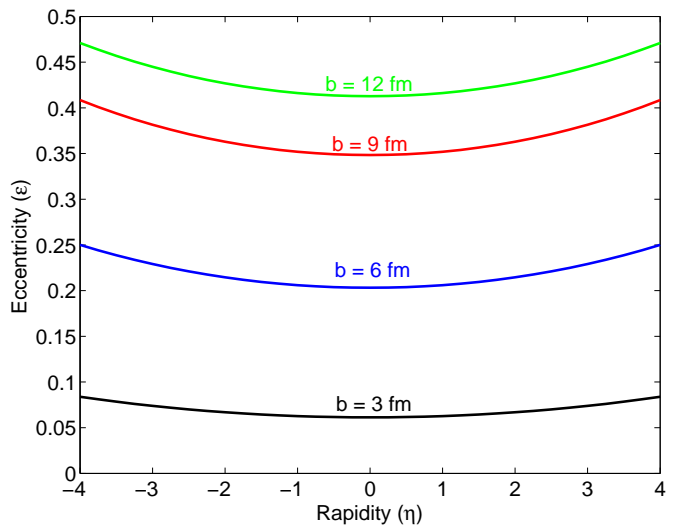,width=3.5in,angle=0}
  \caption{The rapidity dependence of the elliptic
eccentricity of the twisted 
sQGP. (In color online)}
    \label{eccentricity}
\end{figure}

Figures
\ref{contours} and \ref{partdensonline}
 show the 3D diffuse participant geometry of
the twisted  sQGP initial conditions (\ref{dNdxdydeta})
in the transverse plane in several impact parameter and rapidity
slices. For $b=0$ there is complete azimuthal 
symmetry at all rapidities of course.
For finite $b$ the initial sQGP density shifts to the left at forward rapidities and to the right in backward rapidities with our conventions. 
The shifts are small $< 2$ fm,
but distinct. There is also a  small
increase in the eccentricity, $\epsilon(\eta)
=\langle(x-\langle x\rangle)^2-y^2\rangle/\langle(x-\langle x\rangle)^2+y^2\rangle$,
of the twisted sQGP density
away from 
midrapidity as shown in Fig.\ref{eccentricity}.

\section{Azimuthal Dependence of the Twisted sQGP Opacity} \label{number2}

\begin{figure*}[t!]
\begin{center}
%\hspace*{-.6in}
\psfig{file=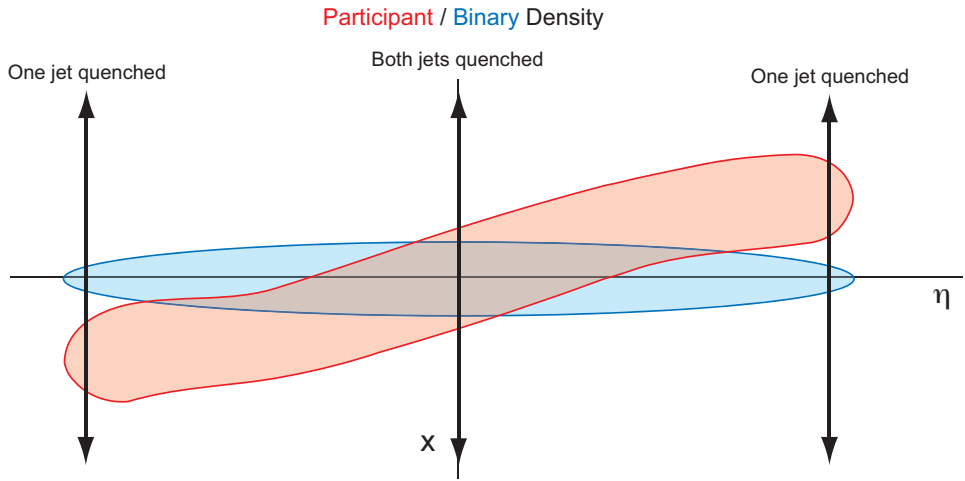,width=5in,angle=0}
\caption{A schematic illustration of the orientation
of the initial twisted sQGP ({\protect{\ref{dNdxdydeta}}}) relative
to the symmetric binary collision distribution ({\protect{\ref{nbin}}}) 
of jet production points projected onto the $(x,\eta)$ reaction plane.
The extra azimuthal asymmetry and octupole twist of jet quenching 
results from the relative rotation of these two distributions. (In color online)
}
\label{binandpart}
\end{center}
\end{figure*}

Unlike the  participant scaling bulk sQGP density,
the binary jet distribution, (\ref{nbin}), does not manifest
a rapidity asymmetry at least at enough high $p_T$
where shadowing is expected to be small.
The jet production points are therefore not
rotated away from the beam direction in this limit.
This is illustrated in Fig.\ref{binandpart}.

Jet quenching due to energy loss in the medium
depends on the opacity and hence density of the plasma
in the direction of propagation\cite{Gyulassy:2003mc,Gyulassy:2000gk}.
The mismatch between the distribution of jet production points and
the rotated distribution of the bulk sQGP matter
therefore induced a peculiar azimuthal dependence
of the weighed opacity
\beqar
&\;& \chi_{\alpha}(x_0,y_0,\eta,\phi,b) = c_\alpha \int_{\tau_{0}}^{\infty}
dt\; t^{\alpha} \nonumber \\ 
&\;& \frac{dN_{g}}
{dxdyd\eta}(x_{0}+t\cos(\phi),y_{0}+t\sin(\phi),b)
\; \; . 
\eeqar{chi}
Different $\alpha=-1,0,1$ weights correspond to
different mechanisms of energy loss.
Elastic scattering energy loss through longitudinally
expanding\cite{Bjorken:1982qr} (or static) sQGP 
matter corresponds to $\alpha=-1\; (0)$.
Inelastic radiative energy loss through 
longitudinally Bjorken expanding (or static) sQGP matter corresponds, on
the other hand, to $\alpha=0\;(1)$.
The results turned out numerically to be
within 10\% in all cases.
Therefore, we drop from now on the $\alpha$ index and show results only for 
$\alpha=0$.

\begin{figure}
\centering
 \epsfig{file=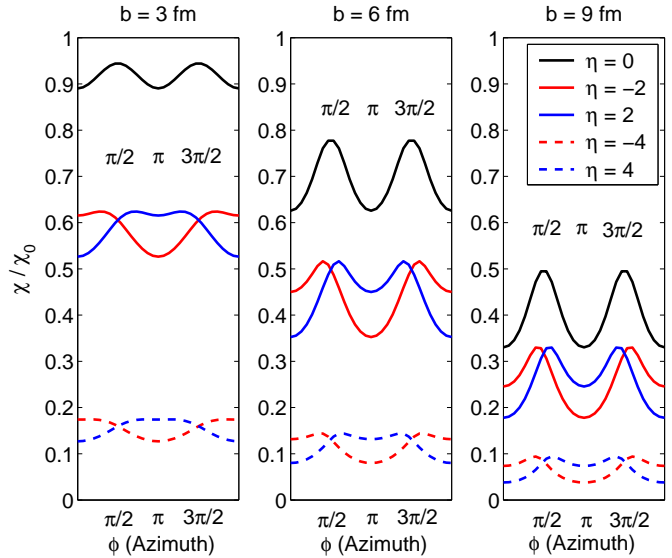,width=3.5in,angle=0}
  \caption{Opacity $\chi$ as a function of azimuthal angle $\phi$ 
for different values of rapidity and impact parameter.  The jet is 
assumed to be born at the origin of the transverse plane.  The
 functions have been normalized to $\chi_{0}$, the opacity felt
 by a jet propagating along the x-axis after being born at the origin
 of the transverse plane, at midrapidity and with impact parameter zero. (In color online)}
    \label{OpacFig}
\end{figure}

\begin{figure}
\centering
 \epsfig{file=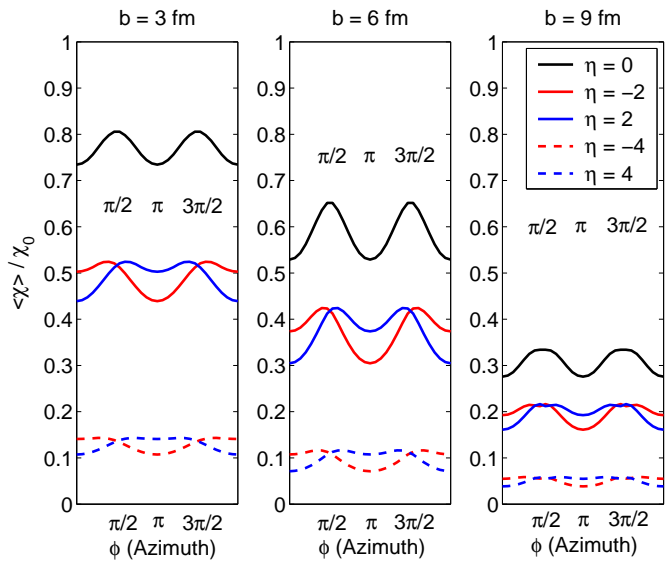,width=3.5in,angle=0}
  \caption{Averaged Opacity $\chi$ as a function of azimuthal angle
 $\phi$ for different values of rapidity and impact parameter. 
 The initial position of of the jet seeing the opacity is averaged 
over the binary distribution.  The functions have been normalized to
 $\chi_{0}$, the opacity felt by a jet propagating along the x-axis 
after being born at the origin of the transverse plane, at midrapidity 
and with impact parameter zero. (In color online) }
    \label{AvOpacFig}
\end{figure}

We shown in Figs. \ref{OpacFig}, \ref{AvOpacFig}
the $\phi$ dependence of the opacity for different
$b$ and $\eta$. In all cases the opacity is normalized to the maximal
$x_0=y_0=b=\eta=0$ opacity
\beq
\chi_0\equiv \chi_0(0,0,0,0)
\eeq{chi0}

In Fig.
\ref{OpacFig}, $\chi_0(0,0,\eta,\phi)/\chi_0$
is shown for a jet produced at the origin
$x=y=0$ for $b=3,6,9$ fm.  
For each $b$, we consider the azimuthal variation
at $\eta=0,\pm 2,\pm 4$.
First, note that the magnitude of the opacity is proportional to the
decreasing central density as either $b$ or $|\eta|$ increases.
Second, the amplitude of the elliptic second moment
$\langle \cos(2\phi)\rangle$ increases as the eccentricity
of the sQGP increases with $b$ as seen in Fig.\ref{eccentricity}.
Third, the maximum opacity at $\eta=0$ is in directions
perpendicular to the impact parameter vector where there is more matter
and minimum parallel to it. 
These features and 
trends are the familiar ones in the conventional standard Bjorken 
scenario\cite{Gyulassy:2000gk}.

However, the new twists caused by the intrinsic longitudinal boost violating
$\delta=O(1)$ initial conditions illustrated in Fig.9
are (1) the emergence of a long range
rapidity asymmetry between $+\eta$ and $-\eta$
with a 180$^0$ phase shifted azimuthal dependence
and (2) an azimuthal pattern 
that has odd harmonics in addition to quadrapole $v_2=cos(2\phi)$ moment.
It is clear that there is a significant high $p_T$ 
$v_1=cos(\phi)$ dipole moment 
away from midrapidity. This is manifest in the difference
between $\chi(\phi=0,\eta)$ and $\chi(\pi,\eta)$
as expected from Fig. 9. 

The most interesting new and subtle 
feature in Fig. 10 , however, is that the azimuthal angles
of maximum opacity
for $\eta\ne 0$
are no longer at $\phi=\pi/2$ and $3\pi/2$, 
as at $\eta=0$, but rotated slightly
away from the normal to the reaction plane.

The splitting of the maxima is the
 unique new signature of the Bjorken violating
initial conditions considered here.
Both $v_1$ and $v_2$ at high rapidities are well known and expected
in nuclear collisions due to directed and elliptic hydrodynamic flow 
of the bulk matter (see Stoecker in ref.2 and \cite{Retiere:2003kf}).
However, the long range rapidity
anti-correlation of the azimuthal rotation of opacity maxima, that as
we show below correspond
to jet quenching minima, is the smoking gun
that we propose as the test of twisted sQGP initial conditions.

To test the robustness of this signal,
we next average over all the initial
jet production points using the binary collision number distribution
\begin{equation}
<\chi> = \frac{1}{dN_{\textrm{Bin}}/d\eta}\int dx_{0}dy_{0}
 \frac{dN_{\textrm{Bin}}}{dxdyd\eta}(x_{0},y_{0},b)\chi
\label{avchi}
\end{equation}
The results shown in  Figure
\ref{AvOpacFig} show all the the same qualitative effects as in the
center jet results in
Fig.\ref{OpacFig} but
slightly diluted due to the extra averaging over $\vx_0$.

In the case that we consider here, the relatively normalized 
$dN_{\textrm{Bin}}$ actually has no
$\eta$ dependence. At moderate $p_T$ the interplay between Cronin and shadowing
could give rise to a $\eta$ dependence that we will report in 
a subsequent paper. 

\section{Azimuthal Dependence of $R_{AA}$} \label{number3}

The observation of jet quenching in different $\eta$ slices
makes it possible to extend
2D jet tomography to 3D. The observable is
the nuclear modification factor
\begin{equation}
R_{AA}=\frac{dN_{AA}/d\eta d^{2}p_{T}}{T_{AA}d\sigma_{pp}/d\eta d^{2}p_{T}}
\label{RAAdef}
\end{equation}
This is the ratio between the spectrum of high $p_T$ hadrons
produced in $A+A$ and the inclusive
differential cross section in
proton-proton collisions  scaled by the binary transverse 
density, $T_{AA}(b)=\int d^2\vx T_A(r_+)T_A(r_-)= N_{bin}(b)/\sigma_H$. 

In general, the computation of $R_{AA}(\eta,\phi,p_\perp))$ 
requires averaging over the power law pQCD parton spectra,
using the distribution of energy loss as in the Gyulassy-Levai-Vitev (GLV) 
or Baier-Dokshitzer-Mueller-Schiff (BDMS) theory\cite{Gyulassy:2003mc}, and integrating over
hadron fragmentation functions. This is beyond the scope
of the present paper. Since our aim here is mainly to introduce and 
illustrate the qualitative features of twisted sQGP initial conditions,
the simple phenomenological 
jet absorption model used in
ref.\cite{Drees:2003zh} will suffice for our purposes.
In this model the survival of a jet in direction $\phi$ relative to the reaction plane is simply assumed to be $\exp(-\kappa \chi(\phi)$.
Averaging over the production points and taking into account
the particular geometric dependence of the line integral
for opacity (\ref{chi}) for our geometry, 
the nuclear modification factor is given by
\begin{equation}
R_{AA}(\eta,\phi;b)=\frac{1}{N_{\textrm{Bin}}} \int
 d^2\vx_{0}\frac{dN_{\textrm{Bin}}}{d^2x}(\vx_{0},b)
 e^{-\kappa\chi(\vx_0,\eta,\phi,b)}
\label{RAAgeom}
\end{equation}
We simplify the computation of $\kappa$
by simply fitting the observed\cite{Adcox:2004mh} 
approximately $p_T$ independent
central collision $R_{AA}\approx 0.2$ in $Au+Au$ at
 $\sqrt{s}=200$ AGeV. (This is achieved if $\kappa \approx 0.25$
in units where the central opacity for $b=0$ is $\chi_0=11/fm^2$).
\begin{figure}
\centering
 \epsfig{file=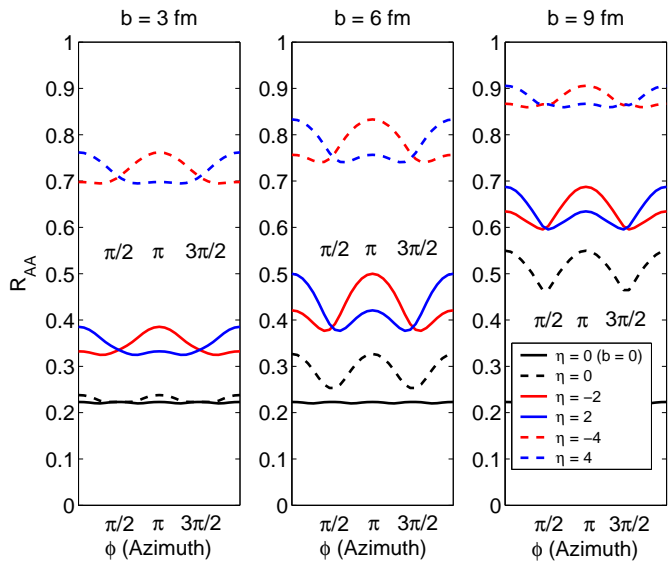,width=3.5in,angle=0}
  \caption{The jet survival probability, $R_{AA}$,
 as a function of azimuthal angle $\phi$ for different 
values of rapidity and impact parameter.  
Note 
the shifted minima of $R_{AA}$ for $\eta=\pm 2$ near 
$\phi=\frac{\pi}{2},\frac{3\pi}{2}$
and the overall $\pi$ phases shift characteristic of the twisted
sQGP initial conditions as in Figs. 10,11. (In color online)}
    \label{RAAFig}
\end{figure}

The computed jet survival probability, $R_{AA}$,
 as a function of azimuthal angle
for different rapidities and impact parameters can be seen in Figure
\ref{RAAFig}.  The solid black line in the figure is the $R_{AA}=0.2$
independent of $\phi$ is 
baseline fit for $\eta,b=0$ . The main features to note are the
same as in Figs. 10,11 but with opacity maxima replaced by
slightly shifted $R_{AA}$ minima.  
First, the magnitude of $R_{AA}$ increases with both increasing
impact
parameter as well as $\eta$.  
Next, note that at
midrapidity, $\eta=0$, there is a symmetry between $\phi = 0$ and $\phi=\pi$
as usual. The elliptic distribution at $b>0$ but $\eta=0$
leads to the usual minimum of the survival probability at 
$\phi=\pi/2,3\pi/2$.

However, the reflection
symmetry is broken away from mid rapidity 
because the jet source and bulk sQGP are shifted relative to each other.
As expected from Figs.10,11, the minimum survival probability
is rotated away from the reaction plane normal
in opposite directions in the forward and backward rapidity
region. To better quantify this effect we introduce in the next section
the octupole twist observable.

\section{The Octupole Twist and Fourier Decomposition} \label{number4}

\begin{figure}
\centering
 \epsfig{file=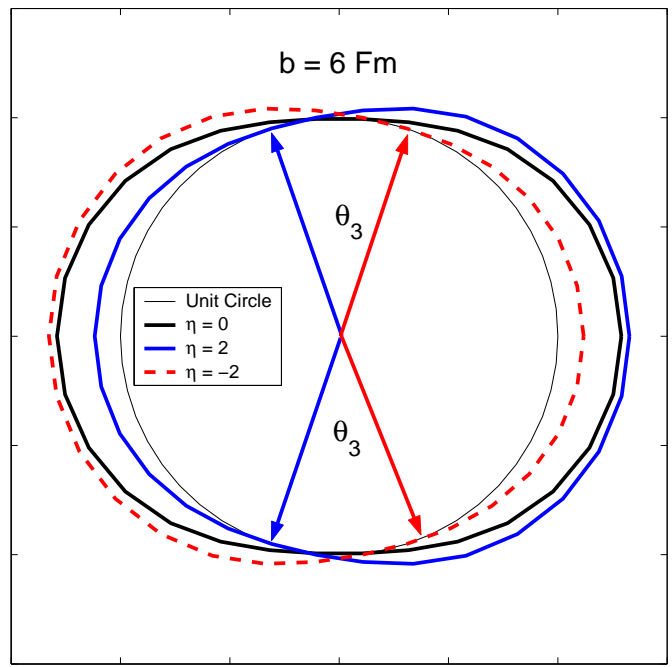,width=3.5in,angle=0}
  \caption{A polar plot of the relatively normalized $R_{AA}/R^{min}_{AA}$
for $b=6$ comparing two  $\eta=\pm 2$ slices (blue solid, dashed red)
to the unit circle corresponding to $b=0$ (thin black).
For $b=6$ but $\eta=0$ (thick black), 
the minima are at $\phi=\frac{\pi}{2},\frac{3\pi}{2}$
and maxima are at $\phi=0,\pi$.
However, for both $b>0$ and $|\eta|>0$,
 twisted sQGP initial conditions
lead to an octupole ``pear'' deformation  with the minima
shifted to $\phi=\frac{\pi}{2}+\frac{\theta_3}{2},
\frac{3\pi}{2}-\frac{\theta_3}{2}$.
The octupole twist angle, $\theta_3(\eta)$, is 
a measure the long range in rapidity
correlations caused by intrinsic local $O(A^{1/3} /Y)$ 
violations of longitudinal
 boost invariance in $A+A$. (In color online)}
    \label{Polar}
\end{figure}

One way to visualize the azimuthal asymmetry 
of $R_{AA}$ is to compare the jet survival probability
in forward and backward rapidity on the same 
polar plot as shown in Figure
\ref{Polar}.  For each fixed rapidity slice and impact parameter, 
we normalize $R_{AA}(\eta,\phi; b)$ by its minimum
$R_{AA}^{min}$ over $\phi$ and plot 
the resulting $R_{AA}/R_{AA}^{min}$ as a function
of $\phi$.

For $b=0$ collisions, this relatively normalized ratio is a unit circle independent
of $\phi$ and $\eta$. It is shown as a thin circle.
For $b>0$, but $\eta=0$ the transverse density profile has an elliptic
shape centered at $x=y=0$ with its 
longest axis oriented normal to
the reaction plane as seen in Fig.9. This leads to the highest survival
probability in the reaction plane $\phi=0,\pi$, as shown by the
thick solid black line in Fig.\ref{Polar}.

For both $b>0 $ and $\eta> 0$, the bulk matter is shifted toward
$\phi=\pi$ (in our conventions) leading to a smaller survival probability than in the $\phi=0$ direction as shown by the solid (blue shifted) pear shape
curve in the figure. The long range
twisted sQGP initial conditions lead on the other hand
for $\eta=-|\eta|<0$ to an  azimuthal pattern  
(red shifted dashed) that
 is simply the reflected version of the $\eta>0$ curve about the $y$ axis.
The pair of minima of both curves are indicated by the arrows. 

We define an ``octupole twist'' angle, $\theta_3(\eta,b)$
to help quantify the magnitude of these 
long range rapidity correlations
as
\begin{equation}
\theta_3(\eta,b)=\phi(R^{min}_{AA}(\eta,b))-\phi(
R^{min}_{AA}(-\eta,b))
\label{rapeq}
\end{equation}
\begin{figure}
\centering
 \epsfig{file=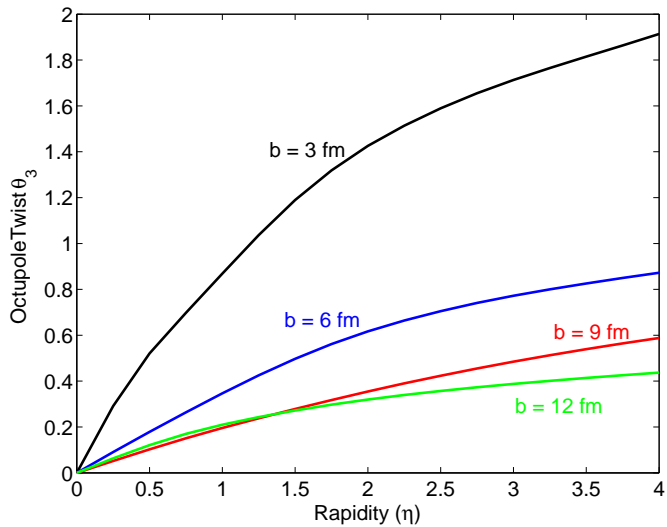,width=3.5in,angle=0}
  \caption{The octupole twist angle $\theta_3(\eta)$ 
of the minimum jet survival direction as in Fig. 13
for different values of the impact parameter are shown. (In color online)}
    \label{rapshift}
\end{figure}
The minima in the forward region are tilted to
$(\pi-\theta_3)/2$ and $(3\pi+\theta_3)/2$
while in the backward region they are tilted in the opposite direction.
\begin{figure}
\centering
 \epsfig{file=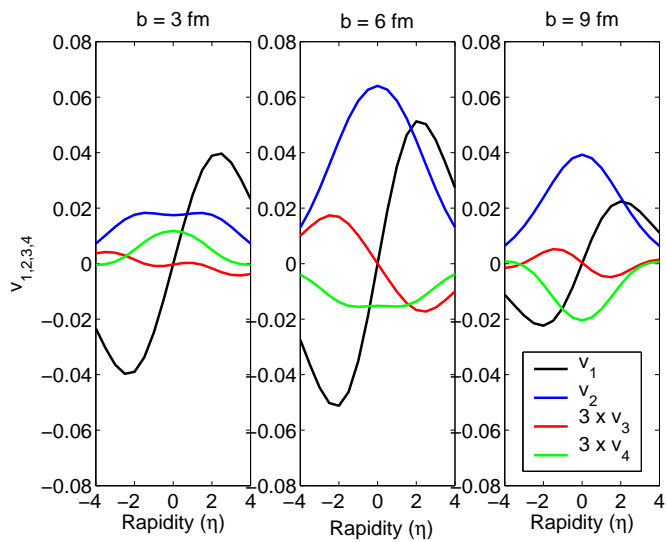,width=3.5in,angle=0}
  \caption{Plots of the Fourier moments of $R_{AA}$ against rapidity 
for different values of the impact parameter.  Note that the third and fourth 
moments are shown multiplied by 3. (In color online)}
    \label{Moments}
\end{figure}

\begin{figure}
\centering
 \epsfig{file=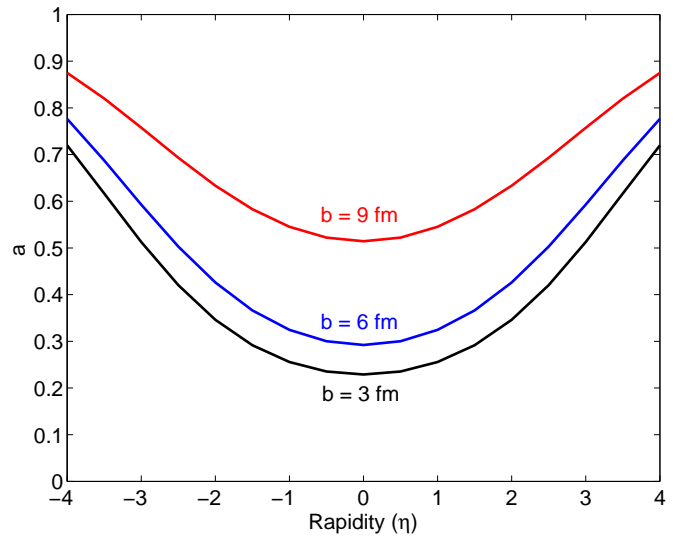,width=3.5in,angle=0}
  \caption{$a(\eta,b)$ versus $\eta$ for different values of the 
impact parameter. (In color online)}
    \label{const}
\end{figure}

Note that the reaction plane orientation can be determined 
unambiguously experimentally via standard low $p_T$ bulk
 directed and elliptic flow $v_1, v_2$\cite{starv2}. 
Jet tomography of high $p_T$ hadrons provides a complementary
short wavelength probe of the reaction dynamics
and can also be decomposed into azimuthal harmonics.
as \begin{equation}
R_{AA}(\eta,\phi,p_\perp;b)=a(1+2\sum_{n}v_{n}\cos(n\phi))
\label{vi}
\end{equation}
The numerical evaluation of the first four moments, $v_n(\eta,b)$,
 are shown
in Figures \ref{Moments} and \ref{const}.  In addition
to the usual $v_1, v_2$ and even $v_4$ moments,
the new feature shown here is the existence of an octupole
$v_{3}=\langle \cos(3 \phi)\rangle $ harmonic
that is responsible for the pear shape
deformation of $R_{AA}$ in Fig. 13.
The magnitude is relatively small though, on the order of $v_4$.
However, $v_4$ for low $p_T$
has already been successfully measured\cite{starv2},
and with a large luminosity upgrade and advanced detectors,
a measurement of $v_3$ at $\eta=\pm2$ may thus
become possible at a future RHIC II\cite{rhicII}.

The predicted peak of the high $p_T$ $|v_1|$ at $\eta\sim 2$,
(with opposite sign relative to the spectator $v_1$)
is another characteristic feature that can be used
to test twisted sQGP initial conditions via 3D jet tomography.

\section{Conclusions} \label{number5}

In this paper we applied the local trapezoidal
multi-particle production BGK model to predict the magnitude
of intrinsic longitudinal boost violating features
of the bulk matter produced in $A+A$ collisions.
We noted that this is a generic feature of both
local mini-jet based dynamics as encoded in LUND string
and HIJING models as well as $k_T$ factorized production models
such as the CGC. The resulting geometrical picture is that 
the sQGP initial conditions has the expected transverse elliptical
shape for non central collisions,  but the bulk density is
twisted away from the beam axis in the reaction ($x,\eta$) plane.

In contrast, the binary collision distribution of hard 
pQCD jet sources remains reflection
symmetric in the transverse plane oriented parallel
to the beam axis at least in the approximation that
Cronin and shadowing can be neglected at high enough $p_T$.
These approximations will be relaxed in a subsequent paper
to explore further how the $p_T$ dependence
3D jet tomography of twisted sQGP initial
conditions can be used to further
test different models of gluon saturation.

The
 generic displacement of the jet sources and  bulk 
matter away from central rapidities was predicted
to produce a novel jet absorption pattern
that deviates from what is now
known at mid-rapidities. The most intriguing is the predicted ``octupole twist''
of the minimum of $R_{AA}$ away from the normal of the reaction plane
and a peak of the jet $v_1$ near $\eta\sim \pm 2$.
These can be further quantified 
in terms of a non-vanishing octupole harmonic of the azimuthal distribution.
Even though the predicted $v_3$ is small, on the order of $v_4$
it may be measurable at  RHIC II \cite{rhicII}. 

The long range in rapidity $\Delta\eta \sim 4-6$ 
correlations induced by  twisted
sQGP initial conditions are expected to persist even at LHC energies.
This is because the dimensionless parameter, $\delta \sim A^{1/3}/\log(s)$,
that controls this phenomenon depends on the logarithm of the collision energy.  Logarithms are very slowly varying functions of the energy and thus the rate of reduction in $\delta$ with energy will be concomitantly slow.

\begin{acknowledgments}
Discussions with J. Harris, T. Hirano, W. Horowitz, M. Lisa,
I. Vitev, and X.N. Wang 
are gratefully acknowledged.
  This work is supported in part by the United States
Department of Energy
under Grants   No. DE-FG02-93ER40764.
\end{acknowledgments}

%\begin{thebibliography}{80}

%\end{thebibliography}

\end{document}